\documentclass[24pt]{article}
\usepackage{amssymb}
\usepackage{epsfig,caption,graphicx,eepic,epic}
\usepackage{pstricks}
\usepackage{amssymb,amsmath}
\usepackage{multido}
\usepackage{array}
\begin{document}
%\draft
\def\be{\begin{equation}}
\def\ee{\end{equation}}
\def\ba{\begin{eqnarray}} 
\def\ea{\end{eqnarray}}
\def\nn{\nonumber}

\newcommand{\bbf}{\mathbf}   
\newcommand{\rrm}{\mathrm}
\title{Open quantum systems at finite temperature\\}

\author{Jean Richert$^{a}$
\footnote{E-mail address: j.mc.richert@gmail.com}\\
and\\
Tarek Khalil$^{b}$
\footnote{E-mail address: tkhalil@ul.edu.lb}\\ 
$^{a}$ Institut de Physique, Universit\'e de Strasbourg,\\
3, rue de l'Universit\'e, 67084 Strasbourg Cedex,\\ 
France\\ 
$^{b}$ Department of Physics, Faculty of Sciences(V),\\
Lebanese University, Nabatieh,
Lebanon}

\date{\today}   
\maketitle 
\begin{abstract}
The consistent definition of the thermodynamic functions of small open quantum systems in contact with an environment in equilibrium with a heat bath has been the subject of many debates in the quantum community.       
In the present work we reproduce and comment parts of a recent approach of this subject by Rivas ~\cite{riv1}. This approach overcomes the controversial discussions generated by the coupling between a system and its environment for any type of coupling between the two parts and allows for a consistent description of the thermodynamical properties of the system strongly interacting with the bath.   
\end{abstract} 
\maketitle
PACS numbers: 03.65.-w, 03.65.Yz, 05.30.-d, 05.70.-a 
\vskip .2cm
Keywords: consistent description of the thermodynamics of coupled open quantum systems.\\

\section{Introduction}
Quantum systems are generally never completely isolated but interact with an environment with which they may exchange energy and other physical observable quantities. Their properties are naturally affected by their coupling to the external world. The understanding and control of the influence of an environment on a given physical system is of crucial importance in different fields of physics and in technological applications and led to a large amount of investigations, see f.i. ~\cite{reb,aba,kos,gol,cor}. 

Under realistic physical conditions the environment will generally show the  properties of a microscopically complex system. If at the time at which the evolution of the system is observed the environment is stationary, i.e. in thermodynamic equilibrium at a fixed temperature $T$,  it is sensible to describe it in the framework of a quantum canonical ensemble if the number of particles it contains is fixed. This is the situation we consider in the present case.

In previous studies ~\cite{kr1,kr2,kr4}  we defined criteria which allow to classify open quantum systems with respect to their behaviour in the presence of an environment at zero temperature. In order to do so we used a general formulation which relies on the examination of the properties of the density operator of the system and its environment. The dynamical behaviour of this operator is governed by the structure of the Hamiltonians of the system, its environment and the coupling Hamiltonian which acts between them. The time-dependent density operator is a crucial quantity in the present context. We work it out for the case where the system is at the temperature of its environment, i.e. staying at thermodynamical equilibrium when evolving in time. 

The introduction of a system in contact with an environment at finite temperature puts the system into a thermodynamical framework. Hence it is described in terms of thermodynamical functions and the description must be consistent with the laws of thermodynamics, in particular the second one concerning the  entropy. This situation is due to the coupling of the system to an external environment. This coupling generates an ambiguity concerning the definition of the system, hence a problem with the determination of the thermodynamic functions and consequently with the rigorous respect of the thermodynamic laws ~\cite{mer,esp1,han,wen,hsi,sei}.

It is the aim of the present dissertation to present and to insist on some concepts which enter the recent work of \'A. Rivas ~\cite{riv1} leading to a rigorous solution of the problem.

The content of the presentation which relies partly on previous work ~\cite{kr5} is the following. In section $2$ we define the system and the environment to which it is coupled at temperature $T=0$ and 
$T > 0$ as well as the expression of the  density matrix which governs the evolution of the system and its environment. In the sequel we shall consider that both the system $S$ and its environment $E$ are the same finite temperature. Section $3$ is the central part of the work where we define first the internal energy in $S$ and the characteristic thermodynamic functions characterizing a stationary system $S$ in equilibrium at the same temperature as $E$, then the same quantities for a system evolving in time, finally the asymptotic behaviour when time goes to infinity as well as the relative entropy evolution which verifies the expected thermodynamical behaviour predicted by the second law. In section $4$ we present the expression of the time dependent density matrix which operates in $S$ space and develop some cases corresponding to different physical situations corresponding to special commutation pproperties of the time independent Hamiltonians which govern the evolution of $S \oplus E$. 
Section $5$ treats the more general case in which $S$ is evolves  under the action of a time-dependent 
Hamiltonian inducing the production or absorption of work. In section $6$ we develop an application of the formalism developed in section $5$ by means of a known model. Finally section $7$ contains a summary, conclusions and possible extensions of the results to other developments. Detailed calculations are developed in appendices. 

\section{The open system}

\subsection{The time independent system and its environment}

Consider a system $S$ and an environment $E$ which acts as a static heat bath, the total system 
$ S \oplus E$ being at thermodynamic equilibrium. The total Hamiltonian reads

\ba
\hat H=\hat H_{S}+\hat H_{E}+\hat H_{SE}
\label{eq1} 
\ea
where $\hat H_{SE}$ defines the interaction between the system $S$ and its environment $E$. We consider here time independent Hamiltonians. 

At temperature $T=0$ the independent systems $S$ and $E$ are described in the bases of states in which they are diagonal

\ba
\hat H_{S}|n\rangle=e_{n}|n\rangle
\notag\\
\hat H_{E}|\nu\rangle=\epsilon_{\nu}|\nu\rangle
\label{eq2} 
\ea

Here $E$ is supposed to be a large system with a fixed number of constituents which exchanges energy with $S$.  If the stationary systems $S$ and $E$ share a common finite temperature $T$ we introduce a canonical description of $S$ and $E$ which takes care of this fact and define the spectral function in terms of the eigenstates $\{\mid \tilde n \rangle\}$ of the effective Hamiltonian $\tilde H_{S}$
and  $\{\mid \tilde \nu \rangle\}$ of the effective Hamiltonian $\tilde H_{E}$.                                                           

The stationary density operators read

\ba
\hat \rho_{S}=\sum_{n}| n\rangle p_{ n}\langle n|=
\sum_{n}|n\rangle \frac{e^{-\beta e_{n}}}{Z_{S}(\beta)}\langle n|
\notag\\
\hat \rho_{E}=\sum_{\nu}|\nu\rangle p_{\nu}\langle \nu|=
\sum_{\nu}|\nu\rangle \frac{e^{-\beta \epsilon_{\nu}}}{Z_{E}(\beta)}\langle\nu|
\label{eq3} 
\ea
where $\beta$ is the inverse temperature $T^{-1}$ and $Z_{S}(\beta)$ and $Z_{E}(\beta)$ the canonical partition functions in both spaces. The Gibbs states in $E$ will be written 

\ba
|\tilde \nu \rangle=\frac{e^{-\beta \epsilon_{\nu}/2}}{{Z_{E}(\beta)}^{1/2}}|\nu \rangle
\label{eq4} 
\ea
and a similar expression in $S$ space

\ba
|\tilde n \rangle=\frac{e^{-\beta \epsilon_{n}/2}}{{Z_{S}(\beta)}^{1/2}}|n \rangle
\label{eq5} 
\ea

In the sequel we shall introduce an {\bf effective} Hamiltonian in $S$ space, $\tilde H_{S}$ which 
has to show and verify two central properties:

\begin{itemize}

\item it must take care of the presence of the coupling to the environment whether weak or strong.

\item it must lead to the determination of the correct thermodynamic functions as mentioned in the introduction.  

\end{itemize}

\subsection{The density operator in $S \oplus E$ space}

The time dependent matrix elements of the density operator $\hat \rho_{SE}(t)$ at $T=0$ are defined as 

\ba
\rho^{k_{1}k_{2}}_{\nu_{3}\nu_{4}}(t)=\sum_{i_{1}i_{2}}\sum_{\nu_{1}\nu_{2}}  a_{i_{1}\nu_{1}}
a^{*}_{i_{2}\nu_{2}}\langle k_{1} \nu_{3}|e^{-it \hat H}| i_{1} \nu_{1}\rangle
\langle  i_{2} \nu_{2}|e^{+it \hat H}| k_{2} \nu_{4} \rangle
\label{eq6} 
\ea
where the $\langle a_{i \nu}\rangle\}$ coefficients are the entanglement amplitudes of the states in $S \oplus E$.
At time $t=0$

\ba
\rho^{\tilde k_{1}\tilde k_{2}}_{\tilde \nu_{3} \tilde \nu{4}}(0)= a_{\tilde k_{1}\tilde \nu_{3}}a^{*}_{\tilde k_{2} \tilde \nu_{4}}
\label{eq7} 
\ea
when the temperature of the system is $T \neq 0$ and the same expression wih the states replaced by the 
corresponding ones given in Eq.2.

Then at $t=0$ with a system $S \oplus E$ for which $S$ and $E$ are decoupled and we introduce f.i.
\ba
\rho^{\tilde k_{1}\tilde k_{2}}_{\tilde \nu_{3}\tilde \nu_{4}}(0)= 
c_{\tilde k_{1}}c^{*}_{\tilde k_{2}}d_{\tilde \nu_{3} \tilde \nu_{4}}
\label{eq8} 
\ea
where the  $d_{\tilde \nu_{3} \tilde \nu_{4}}$ weigh the states 
$|\tilde \nu_{3}\rangle, |\tilde \nu_{4}\rangle$ in $E$ space and the coefficients $c_{\tilde k_{i}}$ weigh
those in $S$ space.

%In the  next section we recall the definitions and derivations of the physical quantities entering the description of an open system coupled to a bath at thermodynamic equilibrium. 

\subsection{Time evolution of the density operator}

In order to obtain the different quantities defined above one needs to determine the density operator in $S$ space $\hat \rho_S(t)$ starting from the expression of the total density operator $\hat \rho_{SE}(t)$ whose matrix elements were written out in section $2.1$. This will be done by means of the projection of 
$\hat \rho_{SE}(t)$ on $S$ space.

A rigorous expression of this operator for any type of interaction $\hat H_{SE}$ and initial conditions 
(i.e. entangled or non-entangled states) can be obtained in whole generality relying on previous work ~\cite{kr5}. Using the same notations as in section $2.1$ it was shown there that the time derivative of the diagonal contributions in $E$ space comes out as 

\ba
\frac{d \hat \rho^{ik}_{S\gamma}(t)}{dt}=(-i)[\hat H_{d}^{\gamma},\hat \rho_{S\gamma}(t)]^{ik}+
(-i)\sum_{\beta \neq \gamma}[\Omega^{ik}_{\gamma \beta}(t)-\Omega^{ik}_{\beta \gamma}(t)]
\label{eq9}
\ea
where $\hat H_{d}^{\gamma}$ is the diagonal part in $E$ space of $\hat H$ for fixed $\gamma$ and  

\ba
\Omega^{ik}_{\gamma \beta}(t)=\sum_{j}\langle \tilde i \tilde \gamma|\hat H_{SE}|\tilde\beta 
\tilde j\rangle 
\langle \tilde j \tilde \beta |\hat \rho_{SE}(t)|\tilde \gamma \tilde k\rangle
\notag\\
\Omega^{ik}_{\beta \gamma}(t)=\sum_{j}\tilde \langle i \tilde \gamma|\hat \rho_{SE}(t)|\tilde \beta 
\tilde j\rangle
\langle \tilde j \tilde \beta |\hat H_{SE}|\tilde \gamma \tilde k\rangle
\label{eq10}
\ea

Integration over time and summation over the $S$ space states $\{\gamma\}$ leads to the matrix elements $\{\rho^{ik}_{S}(t)\}$. The corresponding Gibbs states are obtained in a basis of states in which 
$\hat \rho_{S}(t)$ is diagonal for $i=k$ and by using Eq.(3). The calculation of the matrix elements of $\hat \rho_{S}(t)$  need the use of numerical methods in the general case.

\subsection{Special cases}

The derivation of the previous section Eqs.(9-10) was introduced in order to discuss specific cases of physical interest.

\begin{itemize}

\item $[\hat H_{E},\hat H_{SE}]=0$

In this case Eq.(9) which governs the evolution of $\hat \rho^{ik}_{S\gamma}(t)$ reduces to 

\ba
\frac{d \hat \rho^{ik}_{S\gamma}(t)}{dt}=(-i)[\hat H_{d}^{\gamma},\hat \rho_{S\gamma}(t)]^{ik}
\label{eq11}
\ea
since $\hat H_{SE}$ is diagonal in $E$ in a basis of states in which $\hat H_{E}$ is diagonal. 
It was shown elsewhere that in this case the time evolution of the system is divisible (semi group property of the propagator) like in a Markov process ~\cite {kr3}. One may notice that $S$ and $E$ may be entangled at $t=0$. A model which verifies the considered condition $[\hat H_{E},\hat H_{SE}]=0$ can be found in ref.~\cite{kr5}. This is the  only case for which an entanglement of $S \oplus E$ at $t=0$ induces rigorous divisibility and a Markovian evolution of the system $S$ if the interaction between $S$  and $E$ is stochastic.

\item $[\hat H_{S},\hat H_{SE}]=0$\\

This case corresponds to an adiabatic evolution of the system $S$ which stays in the same state during its time evolution. In a basis of states in which $\hat H_{S}$ is diagonal  

\ba
Tr_{\gamma}\frac {d \hat \rho^{ii}_{S\gamma}(t)}{dt}=
(-i)\{Tr_{\gamma}[\hat H_{S}\oplus\hat I_{E},\hat \rho_{S\gamma}(t)]^{ii}+
[\hat H_{E}\oplus\hat I_{S},\hat \rho_{S\gamma}(t)]^{ii}\}+
\notag\\
(-i)Tr_{\gamma}\sum_{\beta}\{\langle i \gamma|\hat H_{SE}|i \beta \rangle
\langle i \beta|\hat \rho_{S}(t)|i \gamma \rangle-
\langle i \gamma|\hat \rho_{S}(t)|i \beta \rangle
\langle i \beta|\hat H_{SE}|i \gamma \rangle\}
\label{eq12}
\ea
where $\hat I_{S}$ and $\hat I_{E}$ are the unity operators in the corresponding spaces. 

\end{itemize}

\section{Thermodynamical description of the stationary system S}

\subsection{The definition of S in the thermodynamic context: effective Hamiltonian for a time-independent  system S at equilibrium}

Hereafter we follow the method described in ~\cite{riv1,taha} and refs. quoted there. In order to get a consistent description of the system $S$ thermodynamically coupled to its heat bath $E$ one needs to introduce an effective Hamiltonian 

\ba
\tilde H_{S}(\beta)=-\beta^{-1}\ln[\tilde Z_{S} Tr_{E}(\hat\rho_{SE,\beta})]
\label{eq13} 
\ea
where $\tilde Z_{S}=Z_{SE}/Z_{E}$ is an effective partition function, $Z_{SE}$ the partition function of the total system $S \oplus E$ and $\hat\rho_{SE,\beta}$ the corresponding density operator. These definitions leads to an effective density operator 

\ba
\tilde \rho_{S,\beta}=\frac{e^{-\beta \tilde H_{S}}}{\tilde Z_{S}}
\label{eq14} 
\ea

For reasons of consistency the present definitions imply the initial condition for which the density operator of $S \oplus E$ is in product form at $t=0$.

The knowledge of $\tilde H_{S}$ and $\tilde \rho_{S}$ allows the construction of the energy and the thermodynamic functions in $S$ space. The new Hamiltonian $\tilde H_{S}$ is a function of the temperature.
Hence the internal energy $E_{S,\beta}=-\partial_{\beta}ln(\tilde Z_{S}(\beta))$ is given by the expression

\ba
E_{S,\beta}=Tr_{S}[\tilde \rho_{S,\beta}(\tilde H_{S}(\beta)+\beta \partial_{\beta}\tilde H_{S}(\beta))]
\label{eq15} 
\ea

as well as the free energy $\tilde F_{S,\beta}=-\beta^{-1}\ln(\tilde Z_{S}(\beta))$

\ba
\tilde F_{S,\beta}=Tr_{S}[\tilde \rho_{S,\beta)}(\tilde H_{S}(\beta)+ \beta^{-1}\ln(\tilde \rho_{S,\beta}))]
\label{eq16} 
\ea
and the entropy $\tilde \Sigma_{S,\beta}= \beta^{2}  \partial_{\beta}\tilde F_{S,\beta}$

\ba
\tilde \Sigma_{S,\beta}=Tr_{S}[\tilde \rho_{S,\beta}(\beta^{2}\partial_{\beta}\tilde H_{S}(\beta)-\ln \tilde \rho_{S,\beta})]
\label{eq17} 
\ea

The present procedure provides a consistent description of the thermodynamics of open quantum systems. For the explicit derivation of the expressions [15-17] see appendix A.

Relying further on the work of Rivas ~\cite{riv1} we consider next the extension to a time-independent systems $S$ which evolves in time. 

\subsection{Effective Hamiltonian of a system S evolving in time}

In most experimental situations the system $S$ evolves in time. In this case the equilibrium description above does no longer work and one needs to introduce a time-dependence in the description of $S$. There has been a long time of confusion in the search of a correct description of the evolution which warrants a correct definition of the thermodynamic functions and the entropy of an open system.

This goal can be reached if $S$ is governed by an effective Hamiltonian related to $\hat H_{S}$ of the isolated system through the functional relation \cite{riv1,riv2} 

\ba
\tilde H_{S}(t,\beta)=-\beta^{-1}(\hat \rho_{S}(t)- \hat \rho_{S}(0))+\hat H_{S}                                                     
\label{eq18} 
\ea
where $\hat \rho_{S}(t)$ is the density operator which propagates the system $S$ coupled to the bath $E$ through time and, as already mentioned above, consistency implies the specific initial condition 

\ba
\hat\rho_{SE}(0)=\hat\rho_{S}(0)\hat \otimes \hat \rho_{E,\beta}
\label{eq19} 
\ea
where $\hat \rho_{E,\beta}$  is the density operator of the stationary heat bath, i.e. no entanglement between $S$ and $E$ states.

\subsection{Thermodynamical description of the time evolving system S}

Using the defined Hamiltonian $\tilde H_{S}(t,\beta)$ which verifies $\tilde H_{S}(0,\beta)=\tilde H_{S}$
it is easy to derive the expression of the time-dependent internal energy as well as the free energy  

\ba
E_{S,\beta}(t)=Tr_{S}[\hat \rho_{S}(t)[(\tilde H_{S}(t,\beta)+\beta \partial_{\beta}\tilde H_{S}(t,\beta)]]
\label{eq20} 
\ea

\ba
\tilde F_{S,\beta}(t)=Tr_{S}[\hat \rho_{S}(t)(\tilde H_{S}(t,\beta)+ \beta^{-1}\ln(\tilde \rho_{S}(t)))]
\label{eq21} 
\ea
and the entropy of $S$ 

\ba
\tilde \Sigma_{S,\beta}(t)=Tr_{S}[\hat \rho_{S}(t)(\beta^{2}\partial_{\beta}\tilde H_{S}(t,\beta)
-\ln (\tilde \rho_{S}(t)))]                                                            
\label{eq22} 
\ea

These expressions can immediatly be obtained as an extension of the stationary case by introducing the time dependent effective density operator $\hat \rho_{S}(t)$ which will be worked out.

\subsection{Expressions of the thermodynamic functions at infinite time and thermal equilibrium}

At the limit $t= \infty$  and thermal equilibrium these expressions satisfy the standard expressions 

\ba
E_{S,\beta}(\infty)=-\partial_{\beta}\ln Z_{S}=Tr_{S}(\tilde \rho_{S,\beta} \hat H_{S})
\label{eq23} 
\ea

\ba
F_{S,\beta}(\infty)=-\beta^{-1}\ln Z_{S}
\label{eq24} 
\ea

\ba
\Sigma_{S,\beta}(\infty)=-Tr_{S}(\tilde \rho_{S,\beta} \ln \tilde \rho_{S,\beta})
\label{eq25} 
\ea

The expressions are worked out in appendix B.

\subsection{Relative entropy and the second law}

The expressions obtained above lead to a consistent thermodynamical description of the evolution of the system $S$, i.e. they lead to the first and second law. 

In particular, starting from the inequality ~\cite{riv1}

\ba
S [\hat \rho_{S}(t)||\tilde \rho_{S,\beta}(t)] \leq S [\hat \rho_{S}(0)||\tilde \rho_{S,\beta}]                      
\label{eq26} 
\ea
one can derive the thermodynamic inequality 

\ba
S(t)-S(0)-\beta (E_{S,\beta}(t)-E_{S,\beta}(0)) \geq 0                 
\label{eq27} 
\ea

The expressions are worked out in appendix C.

\newpage

\section{The general case - extension to a time dependent Hamiltonian in $S$ space}

We consider now the case for which $\hat H_{S}(t)$ and possibly $\hat H_{SE}(t)$ get time-dependent. It is only case in which work can be generated by the system.

\subsection{Effective time-dependent Hamiltonian} 

The generalization of the Hamiltonian in $S$ space ~\cite{riv1} leads to the expression of the effective Hamiltonian

\ba
\tilde H_{S}(t,\beta)=-\beta^{-1}[\hat \rho_{S}(t,\beta)- \hat \rho_{S}(0,\beta))
\notag\\
-\beta \hat H_{S}(0)-\beta \int_{0}^{t}ds[\hat \rho_{S}(t,\beta)- \hat \rho_{S}(0,\beta)]
d/ds\hat H_{S}(s)]
\label{eq28} 
\ea

\subsection{Internal energy, heat and work}

Using the expression of the corresponding internal energy
 
\ba
E_{int}(t)=Tr_{S}\{\hat \rho_{S}(t,\beta)[\tilde H_{S}(t,\beta)+\beta \partial_{\beta} \tilde H_{S}(t,\beta)\}
\label{eq29} 
\ea 
where the expression of $\partial_{\beta} \tilde H_{S}(t,\beta)$ can be straightforwardly worked out.

It it is easy to verify that

\ba
E_{int}(0)=Tr_{S}\{\hat \rho_{S}(0,\beta)\hat H_{S}(0)\}
\label{eq30} 
\ea
and the heat produced
 
\ba
Q(t)=E_{int}(t)-E_{int}(0)-\int_{0}^{t}ds tr_{S}\hat \rho_{S}(\beta,s)d/ds\hat H_{S}(s)
\label{eq31} 
\ea
where the last term corresponds to work produced by $S$. 

These expressions insure the correctness of the thermodynamic functions.

\section{Application to an example: the Jaynes-Cummings model of interaction between atoms and fields} 

We work out an application of the present formalism with the help of the well known Jaynes Cummings model which describes the evolution of a spinning atom coupled to a single bosonic field.

\ba
\hat H=\hat H_{S}+\hat H_{SE}+\hat H_{E}= \omega_{a}\frac{\hat \sigma_{z}}{2}+\frac{1}{2}\Omega(\hat a \hat\sigma_{+}+ \hat a^{+} \hat\sigma_{-})+\omega_{c}\hat a^{+}\hat a
\label{eq32}
\ea
where $(\hat \sigma_{z}, \hat \sigma_{+}, \hat \sigma_{-})$ are the Pauli matrices, 
$(\hat a, \hat a^{+})$ bosonic annihilation and creation operators, $(\omega_{a}$, $\omega_{c})$ and $\Omega$ energies in units of $\hbar$. 

We consider the particular case for which the spinning atom possesses a time dependent frequency $\omega_{a}(t)$ and the same is supposed to be the case for the coupling strength to the field $\Omega(t)$ 

\ba
\Omega(t)=f(t)\Omega 
\notag\\
f(t)=1-e^{-\alpha t}
\label{eq33}
\ea
where $\alpha$ is a real positive constant.

At $t=0$ the system $S$ (the spinning atom) is decoupled from the environment $E$ (the field) so that the density operator of $S \oplus E$ factorizes into a product state.\\

At each time $|\omega_{c}-\omega_{a}(t)| \ll \omega_{c}+\omega_{a}(t)$. We use the rotating wave approximation by neglecting the contributions of the small period components $(\omega_{c}+\omega_{a}(t))$  and decompose the total Hamiltonian into two parts

\ba
\hat H(t)=\hat H_{1}+\hat H_{2}(t) 
\label{eq34}
\ea
where 

\ba
\hat H_{1}=\omega_{c}(\hat a^{+}\hat a+\frac{\hat \sigma_{z}}{2})
\label{eq35}
\ea
and

\ba
\hat H_{2}(t)= \delta(t)\frac{\hat \sigma_{z}}{2}+\frac{\Omega(t)}{2}(\hat a \hat\sigma_{+}
+ \hat a^{+} \hat\sigma_{-})     
\label{eq36}
\ea 
with $\delta(t)=(\omega_{a}(t)-\omega_{c})$. 

The spinning atom is supposed to possess a spin projection $m=1/2$ and $m=-1/2$ along the quantization axis.

For a fixed eigenstate $\{|n \rangle\}$ The Hamiltonian $\hat H(t)$ can be diagonalized in a $2 \times 2$ basis of states for each state $|n,m\rangle$ where the states $\{|n\rangle\}$ are the eigenstates of $\hat H_{E}$ and $(m=+1/2,-1/2)$ those of $\hat \sigma_{z}$. The eigenvalues read

\ba
E_{n -}(t)=(n+\frac{1}{2})\omega_{c}-\frac{1}{2}[\delta^{2}(t)+(n+1)\Omega^{2}(t)]^{1/2}
\notag\\
E_{n +}(t)=(n+\frac{1}{2})\omega_{c}+\frac{1}{2}[\delta^{2}(t)+(n+1)\Omega^{2}(t)]^{1/2}
\label{eq37}
\ea
In this basis of states the eigenvectors at time $t=0$ are given by

\ba
|\Phi^{n}_{-}(0)\rangle=-\sin(\frac{\alpha_{n}(0)}{2})|n,m=+1/2\rangle+\cos(\frac{\alpha_{n}(0)}{2})|n+1,m=-1/2\rangle
\notag\\
|\Phi^{n}_{+}(0)\rangle=\cos(\frac{\alpha_{n}(0)}{2})|n,m=+1/2\rangle+\sin(\frac{\alpha_{n}(0)}{2})|n+1,m=-1/2\rangle
\label{eq38}
\ea
with $\alpha_{n}(0)=\arctan[\frac{\Omega(0)(n+1)^{1/2}}{\delta(0)}]$. At time $t$

\ba
|\Phi^{n}_{-}(t)\rangle=e^{-i\int_{0}^{t}dt'E_{n -}(t')}|\Phi^{n}_{-}(0)\rangle                                        
\notag\\
|\Phi^{n}_{+}(t)\rangle=e^{-i\int_{0}^{t}dt'E_{n +}(t')}|\Phi^{n}_{+}(0)\rangle
\label{eq39}
\ea
are the eigenstates which are normalized to $1$. The corresponding Gibbs states read

\ba
|\tilde \Phi^{n}_{-}(t)\rangle= d^{1/2}_{nn,-}(t)|\Phi^{n}_{-}(t)\rangle                                       
\notag\\
|\tilde \Phi^{n}_{+}(t)\rangle= d^{1/2}_{nn,+}(t)|\Phi^{n}_{+}(t)\rangle  
\label{eq40} 
\ea
with Gibbs weight factors 

\ba
d_{nn,-}(t)=\frac{e^{-\beta E_{n -}(t)}}
{\sum_{n} e^{-\beta E_{n -}(t)}}                                   
\label{eq41} 
\ea
and a similar expression for $d_{nn,+}(t)$.

It is now possible to construct the density operator in the 2-dimensional $S$ space. The diagonal matrix elements read  

\ba
\rho_{S}^{--}(t,\beta)=\frac{1}{2}\sum_{n}d^{2}_{nn,-}(t)
\notag\\
\rho_{S}^{++}(t,\beta)=\frac{1}{2}\sum_{n}d^{2}_{nn,+}(t)
\label{42}
\ea 
and the non diagonal ones

\ba
\rho_{S}^{-+}(t,\beta)=\frac{1}{2}\sum_{n}d_{nn,-}(t)d_{nn,+}(t)
e^{-i\int_{0}^{t}dt'(E_{n -}(t')-E_{n +}(t'))}
\notag\\
\rho_{S}^{+-}(t,\beta)=\frac{1}{2}\sum_{n}d_{nn,+}(t)d_{nn,-}(t)
e^{-i\int_{0}^{t}dt'(E_{n +}(t')-E_{n -}(t'))}
\label{43}
\ea

The knowledge of the density matrix allows the construction of the effective time dependent Hamiltonian (Eq.28) which acts in $S$ space and last the internal energy of the system (Eq.29).
The expression of the effective Hamiltonian $H_S^{eff}(t)$ acting in $S$ space is developed in Appendix D.

\section{Thermodynamic functions for the case of a time-dependent effective hamiltonian acting in $S$ space} 

\subsection{Internal energy}

Using the explicit expression of the effective Hamiltonian in $S$ space given in Appendix D the internal energy comes out as 

\ba
E_{int}(t,\beta)=\rho_{S}^{+-}(t,\beta)[\tilde H^{-+}_{S}(t,\beta)+\beta \partial\beta\tilde H^{-+}_{S}(t,\beta)]+
\notag\\
\rho^{-+}(t,\beta)[\tilde H^{+-}_{S}(t,\beta)+\beta\partial\beta\tilde H^{+-}_{S}(t,\beta)]+
\notag\\
\rho_{S}^{++}(t,\beta)[\tilde H^{++}_{S}(t,\beta)+\beta \partial\beta\tilde H^{++}_{S}(t,\beta)]+ 
\notag\\                               
\rho_{S}^{--}(t,\beta)[\tilde H^{--}_{S}(t,\beta)+\beta \partial\beta\tilde H^{--}_{S}(t,\beta)] 
\label{eq44} 
\ea

\subsection{Entropy}

The entropy can easily be worked by diagonalization of the density matrix 
$\hat \rho_{S}(t,\beta)$. This leads to the diagonal matrix elements $\rho_{S}^{11}(t,\beta)$ 
and $\rho_{S}^{22}(t,\beta)$. Then

\ba
S(t,\beta)=-[\rho_{S}^{11}(t,\beta)\ln\rho_{S}^{11}(t,\beta)+ 
\rho_{S}^{22}(t,\beta)\ln\rho_{S}^{22}(t,\beta)]+
\notag\\
\beta^{2}[\rho_{S}^{+-}(t,\beta)\partial\beta\tilde H^{-+}_{S}(t,\beta)+
\rho_{S}^{-+}(t,\beta)\partial\beta\tilde H^{-+}_{S}(t,\beta)+
\notag\\
\rho_{S}^{++}(t,\beta)\partial\beta\tilde H^{++}_{S}(t,\beta) 
+\rho_{S}^{--}(t,\beta)\partial\beta\tilde H^{--}_{S}(t,\beta)]
\label{eq45}
\ea

\subsection{Free energy}

Using the expression introduced in Eq.(21) the free energy is  

\ba
F(t,\beta)=\rho_{S}^{+-}(t,\beta)[\tilde H^{-+}_{S}(t,\beta)+
\rho_{S}^{-+}(t,\beta)[\tilde H^{+-}_{S}(t,\beta)+
\notag\\
\rho_{S}^{++}(t,\beta)[\tilde H^{++}_{S}(t,\beta)+
\rho_{S}^{--}(t,\beta)[\tilde H^{--}_{S}(t,\beta)+
\notag\\
\beta^{-1}[\rho_{S}^{11}(t,\beta)\ln\rho_{S}^{11}(t,\beta)+ 
\rho_{S}^{22}(t,\beta)\ln\rho_{S}^{22}(t,\beta)]
\label{46}
\ea

\subsection{Work}

Work produced by the system $S$ is given by the expression in Eq.(31).It reads

\ba
W(t,\beta)=\int_{0}^{t}dsTr_{S}\hat \rho_{S}(\beta,s)d/ds\hat H_{S}(s)
\label{47}
\ea

Using the expressions of the density operator in $S space$ given by  Eq.(42) and the matrix elements of the Hamiltonian in $S$ space in the same basis of states

\ba
H_{S}^{++}(s)=Tr_{E}H_{Sn}^{++}(s)
\notag\\
H_{Sn}^{++}(s)=+1/2\omega_{a}(s)\cos\alpha_{n}(s)
\label{48}
\ea
and
\ba
H_{S}^{--}(s)=Tr_{E}H_{Sn}^{--}(s)
\notag\\
H_{Sn}^{--}(s)=-1/2\omega_{a}(s)\cos\alpha_{n}(s)
\label{49}
\ea
one obtains

\ba
W(t,\beta)=\int_{0}^{t}ds[\rho_{S}^{++}(s,\beta)d/ds H_{S}^{++}(s)+
\rho_{S}^{--}(s,\beta)d/ds H_{S}^{--}(s)]
\label{50}
\ea

\subsection{Heat}

Finally the heat is obtained as

\ba
Q(t,\beta)=E_{int}(t,\beta)+1/2\omega_{a}(0)-W(t,\beta)
\label{51}
\ea

%{\bf A possible verification of the second law: $\Delta S(t)-\beta Q(t) \geq 0$ should be checked..}

\section{Summary and conclusions}

The introduction of thermodynamic concepts in the description of an open quantum system opened many controversial discussions about the consistency of the proposed description with respect to the thermodynamic laws and functions. The difficulties were due to the fact that open systems couple more or less strongly to an environment which generates a major difficulty concerning the definition of the open system itself. These difficulties were overcome very recently ~\cite{riv1,taha} by means of the introduction of an effective temperature dependent and time-dependent Hamiltonian. This led to a correct formulation of the thermodynamic functions and the thermodynamic laws.
Above we developed the essentials of these  approaches. We added an explicit and general description of the reduced density operator which governs the open system under consideration. We discussed the case for which the system and its environment are entangled at the beginning of time in a non-stationary scenario necessarily leading to the lack of divisibility in the time evolution of the system as it is the case if the coupling is of stochastic nature ~\cite{pol1,pol2}. Finally we worked out a specific physical example as an application of the formalism. Further developments can be considered, in particular energy transfer between the system and its environment and the study of thermodynamic cycles.

The present description treats the case of a system and its environment which are maintained at fixed temperature all over the time evolution process. This means that the whole device evolves in a thermodynamic equilibrium process. This may not be necessarily the case in an effective experiment. If stationarity would not be guaranted non-equilibrium processes would enter and hence a non-equilibrium thermodynamic description would have to be introduced.

\section{Appendix A: The thermodynamic functions in the stationary  case}

\begin{itemize}

\item Starting from the canonical definition the internal energy reads 
\ba
E_{S,\beta}=-\partial_{\beta}\ln(\tilde Z_{S}(\beta))=
(-1)(\tilde Z_{S}(\beta))^{-1})\partial_{\beta}\tilde Z_{S}(\beta)
\notag\\
=-Tr_{S}[\tilde \rho_{S,\beta}e^{\beta\tilde H_{S}(\beta)}]\partial_{\beta}\tilde Z_{S}(\beta)
=-Tr_{S}[\tilde \rho_{S,\beta}(-\beta \partial_{\beta}\tilde H_{S}(\beta) -\tilde H_{S}(\beta))                                                                                     e^{\beta\tilde H_{S}}]
\label{eq52} 
\ea
which leads to the expression of eq(15)

\ba
E_{S,\beta}=Tr_{S}[\tilde \rho_{S,\beta}(\tilde H_{S}(\beta)+ \beta\partial_{\beta}\tilde H_{S}(\beta))]                             
\label{eq53} 
\ea

\item The free energy is defined as 

\ba
\tilde F_{S,\beta}=-\beta^{-1}\ln(\tilde Z_{S}(\beta))=
Tr_{S}[\tilde \rho_{S,\beta}(\tilde H_{S}(\beta)+\beta^{-1}\ln \tilde \rho_{S,\beta})]                            
\label{eq54} 
\ea

\item The entropy is related to the free energy through the relation

\ba
\tilde \Sigma_{S,\beta}=\beta^{2}\partial_{\beta}\tilde F_{S,\beta)}=
Tr_{S}[\tilde \rho_{S,\beta}(\beta^{2} \partial_{\beta} \tilde H_{S}(\beta)-\ln\tilde\rho_{S,\beta})]                           
\label{eq55}
\ea

\end{itemize}

\section{Appendix B: Asymptotic behaviour of the thermodynamic functions}

Here we examine the consistence of the definitions of the thermodynamic functions defined above. Under the fixed physical conditions these quantities are aimed to verify strong asymptotic time related conditions ~\cite{riv1}.

\begin{itemize}

\item  At $t=\infty$ the free energy reads

\ba 
\tilde F_{S,\beta}(\infty)=Tr_{S}[\tilde \rho_{S,\beta}(\infty)(\tilde H_{S}(\beta)(\infty)+\beta^{-1}\ln \tilde \rho_{S,\beta}(\infty))]= 
\notag\\
Tr_{S}[\tilde \rho_{S,\beta}(\infty)(\tilde H_{S}(\beta, \infty) +\beta^{-1}\ln (Z_{SE}/Z_{S}Z_{E)}))+
\beta^{-1}\tilde \rho_{S,\beta}(\infty)\ln \tilde \rho_{S,\beta}(\infty)]
\label{eq56}
\ea 
But

\ba 
\tilde H_{S}(\beta, \infty)=\tilde H_{S}(\beta)+\beta^{-1}\ln (Z_{SE}/Z_{S}Z_{E)})
\label{eq57}
\ea 

Simplifying this expression

\ba 
\tilde F_{S,\beta}(\infty)=-\beta^{-1}Tr_{S}[\tilde \rho_{S,\beta}(\infty) \ln (Z_{S}]=
-\beta^{-1}\ln (Z_{S})
\label{eq58}
\ea
and consequently

\ba 
\tilde Z_{S}(\infty)=Tr_{S}[e^{-\beta\tilde H_{S}(\beta, \infty)}]= Tr_{S}[e^{-\beta\hat H_{S}}]=                 Z_{S}
\label{eq59}
\ea

\item  At $t=\infty$

\ba 
\Sigma_{S}(\infty)=\beta^{2}\tilde Z_{S}(\infty)=\beta^{2}\ln Z_{S}
\label{eq60}
\ea
which is a consistent result. 

\end{itemize}

\section{Appendix C: Relative entropy}

The expected inequality given in Eq.(26) relies on the general expression 

\ba
S(\hat \rho_{1}|| \hat \rho_{2})=Tr(\hat \rho_{1}\ln\hat \rho_{1}-\hat \rho_{1} \ln \hat \rho_{2})                       
\label{eq61}
\ea

in Eq.(26) one works the lhs of the inequality

\ba
Tr_{S}[\hat \rho_{S}(t)\ln \hat \rho_{S}(t)]-Tr_{S}[\hat \rho_{S}(t)\ln \hat U_{S}(t)\hat \rho_{S,\beta}
=Tr_{S}[\hat \rho_{S}(t)\ln e^{-\beta \tilde H_{S}(\beta,t)}]
\notag\\
=Tr_{S}[\hat \rho_{S}(t)\ln \hat \rho_{S}(t)]+\beta [Tr_{S}\hat \rho_{S}(t)\tilde H_{S}(\beta,t)] 
\label{eq62}
\ea
where $\hat U_{S}(t)$ is the time propagator of $\hat \rho_{S,\beta}$.

Subtracting a term $\beta^{2}Tr_{S} \hat \rho_{S}(t)\ln \hat \rho_{S}(t)$ from the first term and adding it to the second one leads to the lhs of Eq.(22)

\ba
S [\hat \rho_{S}(0)||\tilde \rho_{S,\beta}(t)] =-S(t)+\beta E_{S,\beta}(t)
\label{eq63}
\ea

The rhs reads
\ba
S [\hat \rho_{S}(0)||\tilde \rho_{S,\beta}]=Tr_{S}[ \hat \rho_{S}(0)\ln \hat \rho_{S}(0)                           
- \hat \rho_{S}(0)\ln \hat \rho_{S,_\beta}]
\label{eq64}
\ea
Subtracting and adding $ \beta^{2}\partial_{\beta}\tilde H(\beta,0)$ on the rhs of Eq.(42) leads to 

\ba
S [\hat \rho_{S}(0)||\tilde \rho_{S,\beta}]=-S(0)+\beta \hat H_{S}(0)  
\label{eq65}
\ea

and grouping the expressions together one finally gets 

\ba
S(t)-S(0)-\beta (E_{S,\beta}(t)- E_{S,\beta}(0)) \geq 0
\label{eq66}
\ea
which proves the consistency of the thermodynamic description of the open system $S$.

\section{Appendix D: Time-dependent effective Hamiltonian acting in $S$ space}

Using the expressions of the matrix elements of the density operator (Eqs.42-43 in section 5) and the conditions imposed to the system at $t=0$ the matrix elements of the effective Hamiltonian acting in $S$ space in the framework of the Jaynes-Cummings model can be put into the following form

\ba
\tilde H^{--}_{S}(t,\beta)=-1/2 \omega_{a}(t)
\notag\\
\tilde H^{++}_{S}(t,\beta)=+1/2 \omega_{a}(t)
\label{eq67}
\ea
and 

\ba
\tilde H^{-+}_{S}(t,\beta)=-\beta^{-1}[\rho^{-+}(t,\beta)-\rho^{-+}(0,\beta)
\notag\\
-\beta\int_{0}^{t}(\rho^{-+}(s,\beta)- \rho^{-+}(0,\beta))d/ds \hat H_{S}^{++}(s)ds]
\label{eq68}
\ea
where $\hat H_{S}(s)]=1/2\omega(s)$ and a similar expression for $ \tilde H^{+-}_{S}(t,\beta)$

\ba
\tilde H^{+-}_{S}(t,\beta)=-\beta^{-1}[\rho^{+-}(t,\beta)-\rho^{+-}(0,\beta)
\notag\\
-\beta\int_{0}^{t}(\rho^{+-}(s,\beta)- \rho^{+-}(0,\beta))d/ds \hat H_{S}^{--}(s)ds]
\label{eq69}
\ea

%{\bf Check the bibliography, add citations of the list at  hand if opportune}

\end{document}